\documentclass{phbauth}        
\usepackage[dvips]{graphicx}

\begin{document}

\begin{frontmatter}

\title{Some low-temperature properties of a generalized Hubbard model
with correlated hopping}

\author{L.~Didukh},
\author{V.~Hankevych\thanksref{thank1}},
\author{Yu.~Skorenkyy}

\address{Ternopil State Technical University, Department of 
Physics, 56 Rus'ka Str., Ternopil UA--46001, Ukraine}
\thanks[thank1]{Corresponding author. E-mail: vaha@tu.edu.te.ua} 

\begin{abstract}
In the present paper we study some correlation effects in a generalized
Hubbard model with correlated hopping within low-temperature region using
a generalized mean-field approximation. 
It is shown that in a series of cases
the model leads to consequences deviating essentially from those 
of the Hubbard model.
We consider the possibility of applying the result to interpret
the peculiarities of physical
properties of systems with narrow energy bands.
\end{abstract}


\end{frontmatter}

Theoretical analyses, on the one hand, and 
available experimental data, on the other hand, point out the necessity of 
the Hubbard model generalization by taking into account correlated 
hopping~\cite{did1,hir1}. 
In the recent few years, similar models 
have been studied intensively (e.g., see Refs.~\cite{amad}
and references therein). In particular, some of
these models~\cite{ex_sol} have been solved exactly in a wide range of 
parameters.

The present paper is devoted to study of correlation effects in a 
generalized Hubbard model with correlated hopping within low-temperature 
region in the case of partially filled narrow energy bands. 

We start from the generalization of the Hubbard model
with correlated hoppings $T_1$ and $T_2$ \cite{did1}:
\begin{eqnarray} \label{ham}
H=&&-\mu \sum_{i\sigma}a_{i\sigma}^{+}a_{i\sigma}+
t(n){\sum \limits_{ij\sigma}}'a_{i\sigma}^{+}a_{j\sigma}+
\\
&&T_2{\sum \limits_{ij\sigma}}' \left(a_{i\sigma}^{+}a_{j\sigma}n_{i\bar 
\sigma}+h.c.\right)
+U \sum_{i}n_{i\uparrow}n_{i\downarrow},
\nonumber
\end{eqnarray}
where all the notations are usual, \mbox{$t(n)=t_0+nT_1$,} $t_0$ is the 
nearest-neighbor hopping integral,
$n$ is the electron concentration, the prime at the sums signifies that 
$i\neq{j}$.

The peculiarity of model~(\ref{ham}) is taking into account 
the correlated hopping $T_1$ leading to the concentration dependence of the
hopping integral $t(n)$ in contrast to similar models 
which are considered in Refs.~\cite{amad}.

In Ref.~\cite{d_h1} we have obtained the single-particle Green function
and the quasiparticle energy spectrum (which has the exact atomic and 
band limits) for arbitrary electron concentration using a 
variant~\cite{did3} of the generalized mean-field approximation~\cite{appr}. 
For the case $n=1$ and $t(n)=-T_2$ the approach recovers (see 
Ref.~\cite{d_h2}) the exact results~\cite{ex_sol}.

The energy gap $\Delta E$ (which is calculated from the spectrum) is 
found to increase with a deviation from half-filling ($\Delta E$ has a 
minimum at $n=1$).  This explains the  fact  that  in
a metallic  phase  the vanadium  oxides  with  fractional  number of
electrons per cation (V$_k$O$_{2k-1}$, $k\geq 3$) exhibit ``metallicity'' to
lesser  extent  than  the compounds VO$_2$  and  V$_2$O$_3$ (integer number 
of electrons per site)~\cite{kos}.

Within low-temperature
region for \mbox{$U\to\infty$} width of the lower Hubbard band at $n<1$ is
$W_l=2w\left[2/(2-n)-n\right]$;
and at $n>1$ width of the upper Hubbard band is
$W_u=2\tilde{w}(n-2+2/n)$,
where $w=z|t(n)|,\ \tilde{w}=z|\tilde{t}(n)|,\ \tilde{t}(n)=t(n)+2T_2,\ z$ 
is the number of nearest neighbors to a site.

\begin{figure}[htbp]
\begin{center}\leavevmode
\includegraphics[width=0.8\linewidth]{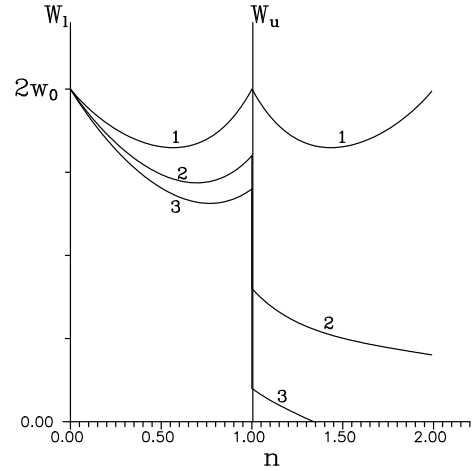}
\caption{
Width of the lower ($W_l$) and upper ($W_u$) bands as a
function of electron concentration $n$: 
1) $\tau_1=\tau_2=0$ (the Hubbard model); 2) $\tau_1=\tau_2=0.2$; 
3) $\tau_1=\tau_2=0.3$.
}\label{fig1}\end{center}\end{figure}

The concentration dependence of subband widths plotted in Fig.~\ref{fig1} is 
caused firstly, by correlation effect of narrowing of subbands,
secondly, by the concentration dependence of the hopping integrals in
the lower $t(n)$ and upper $\tilde{t}(n)$ Hubbard bands (this is a 
peculiarity of the present model). Taking into account correlated hopping
leads to the essential narrowing of subband, and what is more, narrowing by
this factor increases with increasing of electron concentration. 
Fig.~\ref{fig1}
illustrates the non-equivalence of the cases $n<1$ and $n>1$ in the 
framework of the present model (in contrast to the electron-hole symmetry
of the Hubbard model). One can see that at some values of the parameters
$\tau_1=T_1/|t_0|,\ \tau_2=T_2/|t_0|$ correlated hopping
$W_l\gg W_u$, thus effective mass of current carriers within
the lower Hubbard band $m_l\ll m_u$ ($m_u$ is the effective mass of current 
carriers within the upper Hubbard band). So, the narrow-band materials
with conductivity determinated by current carriers within the lower
band can have much smaller resistivity than the compounds with
conductivity determinated by current carriers within the upper band.
This fact is confirmed by the experimental data~\cite{jon}: 
the metalooxides with less than half-filled $3d$-shell (Mn$_2$O) exhibit
much higher conductivity than the compounds with half or more than 
half-filled  $3d$-shell (MnO, NiO).

V.H. thanks the Organizing Committee of the LT22 Conference for a grant.
The authors are grateful to Prof.~D.~Khomskii for valuable discussions on
the part of the result considered in the present work.


\end{document}